\documentstyle[aps]{revtex}

\begin{document}
\twocolumn
\title{ Bosonization in $d=2$ from finite chiral determinants with a Gauss decomposition }

\vspace{.7cm}

\author{
        A N Theron$^{a,b}$, F G Scholtz$^{b}$ and H B
        Geyer$^{b,c}$\cite{AvH}
        }

\address{
          $^{a}$Theory Division, CERN, CH-1211 Geneve 23, Switzerland\\
         $^b$Institute of Theoretical Physics,
         University of Stellenbosch, 7600 Stellenbosch, South Africa\\
         $^c$FB Physik, Universit\"at Siegen, Postfach 101240, D-57068
         Siegen, Germany
}
\maketitle

\begin{abstract}
We show how to bosonize two-dimensional non-abelian models using finite chiral determinants  calculated from a Gauss decomposition.  The calculation is quite
straightforward and hardly more involved than for the abelian case.
In particular,  the counterterm $A\bar A$, which
is normally motivated from gauge invariance and then added by hand,
appears naturally in this approach.
\end{abstract}


\bigskip

\narrowtext



The path integral approach to bosonization of two-dimensional field
theories was developed some time ago \cite{schaposnik} by exploiting
the fact that gauge fields may be decoupled from the fermions by
making local chiral transformations.  It is crucial in this approach
that the anomalous contributions from the fermionic measure under
chiral rotations are properly taken into account.

More recently an approach to bosonization with path integrals, refered
to as smooth bosonization, was introduced \cite{damgaard}.  Here a
chiral gauge symmetry is first introduced, and the bosonization rules
are then obtained by choosing an appropriate gauge.  Again the
anomalous contribution from the path integral measure is of great
importance.  For an infinitesimal chiral transformation this
contribution may be calculated by the method of Fujikawa
\cite{fujikawa}, but for the applications mentioned above it is
necessary to obtain the anomalous Jacobian for a finite chiral
transformation.  One therefore requires the iteration of the
infinitesimal anomaly to obtain the Jacobian for a finite chiral
change of variables.

For the abelian case this problem is rather simple, and the abelian
chiral anomaly is easily integrated to yield the free scalar boson
action \cite{schaposnik}.  The non-abelian case is, however,
complicated and it is difficult to show directly that the non-abelian
anomaly can be integrated to yield the WZW \cite{witten} action.  In
this paper we present a simple and direct derivation of this result
for su(2), hardly more involved than the abelian case.

Although it is well known that the finite chiral transformation yields
the WZW action (or the gauged WZW action in the presence of a
background field) \cite{polyakov}, this result is of such importance
in the path integral bosonization program that we consider it
worthwhile to present an alternative derivation.  The present
derivation is not only more directly constructive than the
conventional one, which proceeds indirectly by showing that the
variation of the WZW model is in agreement with the anomaly
\cite{polyakov}, but it also serves to clarify some issues connected
with regularization.  In this context it is often stated that it is
necessary to introduce a counter term of the type $A\bar{A}$ by hand
as to ensure gauge invariance \cite{polyakov,kaa}.  We show here that
this term arises naturally when the covariant form of the anomaly is
iterated.

The Jacobian associated with the final chiral transformation is
calculated by using the Gauss decomposition for the group.  The
calculation is then relatively simple and the iteration of the
infinitesimal result is hardly more involved than the abelian case.
Normally there is very little advantage in writing the WZW model using
an explicit parameterization for the group, because the action is
complicated and the global symmetries are obscured, but in the Gauss
decomposition the WZW model assumes a remarkably simple form which is
closely related to the Wakimoto free field realization \cite{wakimoto}
of Kac-Moody algebras, as initially noted in Ref.\ \cite{gerasimov}.

We start with the current-current generating functional of free Dirac
fermions,
\begin{equation} \label{5.17}
Z_{\rm F}[A ] = \int \!  D {\bar \psi}
D \psi e^{-\int \! d^2 x\,[ \psi_{1}^{\dagger} i \partial \psi_{1} +
\psi_{2}^{\dagger} i {\bar \partial} \psi_{2} -\psi_{1}^{\dagger} A
\psi_{1} - \psi_{2}^{\dagger} {\bar A} \psi_{2} ] } \; ,
\end{equation}
where $A= A^{a}T^{a}$, $T^a=\frac{1}{2}\sigma^a$, and $\sigma^a$ are
the Pauli spin matrices.  The bar notation is used to denote the
anti-holomorphic components.  The source $A$ for the currents may also
be viewed as a background gauge field.

We restrict ourselves to two fermion colours only, so that we can view
the fermions to be in the fundamental representation of SU(2).
Actually we have a U(2) symmetry, but we shall focus attention on the
su(2) subalgebra.

In many applications it is possible to choose the light cone gauge, so
that only one component of the vector $A$ is present, but we shall be
more general for the moment.

We find it convenient to  introduce the Cartan-Weyl basis,
\begin{eqnarray} \label{n.3}
T^+ &=&T^1+iT^2\nonumber\\
T^- &=&T^1-iT^2\\
T^0&=&T^3\nonumber
\; ,
\end{eqnarray}
so that we are in fact considering the complexification of su(2).  In
this basis we denote the components of the gauge fields by $A^\pm$,
$A^0$.  It has been noted \cite{rothe} that it is convenient to
consider the complexified group when considering path integral
bosonization, and the basis (\ref{n.3}) is useful in establishing a
connection with the Wakimoto realization.

We now calculate the anomaly resulting from the following change
of variables on the fermions,
\begin{eqnarray} \label{n.1}
\psi_1 &\rightarrow& \psi^{\prime}_1=g\psi_1\,,\quad
\psi_1^{\dagger} \rightarrow \psi^{\dagger\prime}_1
=\psi_1^{\dagger}g^{-1}, \nonumber  \\
\psi_2 &\rightarrow& \psi^{\prime}_2 =\psi_2\,,\quad
\psi_2^{\dagger} \rightarrow \psi^{ \dagger
\prime}_2=\psi_2^{\dagger}. \end{eqnarray}
If $g$ is infinitesimal, $g(\epsilon)=e^{i\epsilon T^a \theta^a}$, and
we have the familiar  anomalous contribution
\begin{equation} \label{n.2}
J_\epsilon=\frac{\epsilon}{\pi}\int d^2x \rm{Tr(\theta  F)},
\end{equation}
with $ F=\epsilon_{\mu \nu}F^{\mu \nu}$ the dual field strength and
$\theta =\theta^a T^a$.  The Jacobian (\ref{n.2}) originates from the
non-invariance of the path integral measure under the chiral rotation,
as is most naturally established using the methods of Fujikawa
\cite{fujikawa}.  We prefer the above covariant form of the anomaly,
as its iterated finite form will be the gauged WZW model.

When making the finite rotation (\ref{n.1}), we change the external
gauge field used in regulating the measure, so that the anomaly
associated with the finite rotation is given by
\begin{eqnarray}
\label{n.4}
J=\frac{1}{\pi}\int\!d^2x &&\int_0^1\!\!\!d\epsilon
{\rm Tr }\{i\theta  {\bar \partial}
(g^{-1}(\epsilon)A g(\epsilon)
- g^{-1}(\epsilon)i \partial g(\epsilon))\nonumber \\
&&-\partial {\bar A}-[(g^{-1}(\epsilon)A g(\epsilon)
- g^{-1}(\epsilon)i \partial g(\epsilon),{\bar A}]\}.
\end{eqnarray}
This expression is nothing but the gauged WZW-action
\begin{eqnarray}
 \label{gaugewzw}
  S_{\rm WZW}= - S[g] &-&           \frac{i}{4\pi}\int \!  d^2 x \, [
{\rm Tr}(A(\partial_{-} g)g^{-1}) \nonumber \\ &+&
{\rm Tr}(\bar Ag^{-1}\partial_{+} g)  + AgA_{-}g^{-1}    + A\bar A]\,,
 \end{eqnarray}
with \begin{eqnarray}
\label{4.10}
  S[g] &=&  \frac{1}{8\pi}\int\!d^2 x\,{\rm Tr}(g \partial_{\mu}
         g^{-1} g\partial_{\mu} g^{-1})
        \nonumber \\
       & &
       +\frac{1}{12\pi}\int_{\Gamma}\!d^3 x\, \epsilon_{\mu \nu \rho}
       {\rm Tr}(g\partial_{\mu} g^{-1}g\partial_{\nu}g^{-1}
       g\partial_{\rho} g^{-1})\; ,
\end{eqnarray}
as we demonstrate explicitly below.

If we would have worked in the light cone gauge ${\bar A}=0$,
very often the prefered gauge when two-dimensional gauge theories are
studied, we would have obtained instead
\begin{equation}
\label{n.4a}
J=\frac{1}{\pi}\int d^2 x \int_0^1 d \epsilon
{\rm Tr }\{i\theta  {\bar \partial}
(g^{-1}(\epsilon)A g(\epsilon))\}
\; .
\end{equation}
This action is expression (\ref{gaugewzw}) with ${\bar A}=0$.  For
transparency the discussion below is restricted to this case.  Results
for the general case follow without any complications, as we also
elaborate below.

A direct evaluation of the epsilon integration in (\ref{n.4}) is
generally not feasible.  However, by first introducing the Gauss
decomposition for SU(2) these integrations can easily be done.  We
therefore write for the group element $g$
\begin{equation}
\label{n.5}
g=e^{ iT^+ \theta^-   }  e^{i T^0 \theta^0}e^{i T^- \theta^+}
\; .
\end{equation}
For a general element of SL(2,C) the $\theta^a$ are all complex,
but for the  identification of the Wakimoto realization we
consider the transformation (\ref{n.5}) with $\theta^0$ real and
$\theta^-$ the complex conjugate of $\theta^+$.

The transformation (\ref{n.1}) is now performed in three steps.  First
we make a $T^+ $ rotation, then a $T^0$ rotation, and finally a $T^-$
transformation.  We therefore start by performing the local chiral
transformation
\begin{eqnarray} \label{5.19}
\psi_1 & \rightarrow  e^{iT^+\theta^-}\psi_1 \; ,
\; \;   \psi_2 & \rightarrow  \psi_2 \; , \\
\psi^{\dagger}_1 & \rightarrow \psi^{\dagger}_1
e^{-i T^+\theta^-}\; ,\;\;
 \psi^{\dagger}_2 & \rightarrow \psi^{\dagger}_2
\; .
 \end{eqnarray}
The Lagrangian  of  eq.\ (\ref{5.17}) (with $\bar A=0$) changes to
\begin{equation}
\label{5.21}
{\cal L}\rightarrow  {\cal L}_{\rm F}-\psi^{\dagger}_1  A^{\prime}
\psi_1
\end{equation}
under transformation (\ref{5.19}). Here  ${\cal L}_{\rm F}$ is the Lagrange
density for free Dirac fermions and
\begin{eqnarray} \label{5.22}
A^{\prime}
& = & -\partial \theta^-T^+ +
A^-T^+ +A^0(T^0+iT^+\theta^-)
\nonumber\\
&&
+A^+(T^- -2i\theta^-T^0+(\theta^-)^2 T^+)
\; .
\end{eqnarray}
For the transformation
\begin{equation}\label{5.25}
   \psi_1\rightarrow e^{iT^+\theta^-}\psi_1
\end{equation}
the contribution from the measure is obtained by substituting
$g(\epsilon)$ by $e^{i\epsilon T^+\theta^-}$ in expression
(\ref{n.4a}).  After evaluating the trace we have the following
anomalous contribution to the action
\begin{equation}\label{5.24}
S_+=\frac{ 1   }{ \pi   }  \int d^2x    \theta^-{\bar \partial
   }A^+\ \; .
\end{equation}
Terms that depend on $\epsilon $ vanish when the
trace is taken, so that the epsilon integration yields one.

The second step is to perform the change of variables
\begin{eqnarray}  \label{5.26}
\psi_-  &\rightarrow &e^{iT^0\theta^0}\psi_-\; ,\nonumber\\
\psi^{\dagger}_-&\rightarrow &\psi^{\dagger}_-e^{-iT^0\theta^0}
\; .
\end{eqnarray}
The Lagrangian density now becomes
\begin{equation}\label{5.28}
{\cal L}\rightarrow {\cal L}_{\rm F}
-\psi^{\dagger}_1  A^{\prime \prime}
\psi_1\; ,
\end{equation}
where
\begin{eqnarray}
\label{5.27}
  A^{\prime\prime}  &=& \partial\theta^0 T^0 -
 (\partial\theta^- T^+)e^{-i\theta^0}
\nonumber\\
&&-[(A^- + i \theta^-A^0+ A^+(\theta^-)^2)T^+e^{-i\theta^0}\nonumber\\
&&+ A^0T^0-2i\theta^-A^+T^0+A^+T^-e^{i\theta^0}]
\; .
\end{eqnarray}

For this transformation the contribution from the measure may again be
evaluated from (\ref{n.4a}) by taking $g(\epsilon)=e^{i \epsilon
\theta^0 T^0 }$, but $A$ has to be replaced by the $A^{\prime}$ of
eq.\ (\ref{5.22}).  The result after the trace has been performed is
the additional contribution
\begin{eqnarray}\label{5.31}
S_0&=&\frac{  1  }{ 2 \pi   }\int d^2x
 \int_0^1 d\epsilon   \theta^0       {\bar
  \partial  } \{ \epsilon    \partial \theta^0 +
 {\bar \partial } A^0-2i\theta^-A^+ \}
\nonumber \\
&=&
\frac{1}{2 \pi}\int d^2x
\{ \frac{1}{2}  \theta^0       {\bar
  \partial  }  \partial \theta^0 +
\theta^0 {\bar \partial } A^0-2i\theta^0\theta^-A^+ \}
\; .
\end{eqnarray}

Finally the change of variables
\begin{equation}\label{5.31.a}
\psi_-\rightarrow e^{iT^-\theta^+}\psi_- \; ,
\end{equation}
is performed.  We repeat the above step for $T^-$, taking into account
that the background field $A$ has now been altered by both the $T^+$
and $T^0$ rotations, and that $A^{\prime \prime}$ of expression
(\ref{5.27}) has to be used instead.  This gives the final
contribution
\begin{eqnarray}
\label{5.33}
S_- && =\frac{  1  }{\pi    }\int d^2x \left[\theta^+{\bar     \partial
}\left(\partial\theta^-e^{-i\theta^0} \right. \right.
\nonumber \\
&& \left. \left.    \quad + e^{-i\theta^0}
(A^- + i \theta^-A^0+ A^+(\theta^-)^2)
\right)\right]
\; .
\end{eqnarray}
As for the $T^+$ transformation the integration over
$\epsilon $ is trivial. The  Lagrangian is now
\begin{equation}
\label{5.34}
{\cal L}={\cal L}_{\rm F}+\psi^{\dagger}_1 (g^{-1}i\partial g )\psi_1
 - \psi^{\dagger}_1g^{-1}A  g  \psi_1
\; ,
\end{equation}
where $g$ was given in (\ref{n.5})

The total anomalous contribution from the measure to the action
follows from expressions (\ref{5.24}), (\ref{5.31}) and (\ref{5.33})
as
\begin{eqnarray}
\label{5.36}
S &=&\frac{ 1  }{ \pi   }\int d^2 x
[\frac{  1  }{ 4   }\partial \theta^0{\bar
  \partial  }\theta^0 +  {\bar      \partial } \theta^+e^{-i\theta^0}
\partial \theta^-
\nonumber
\\
&&+  A^+({\bar       \partial }\theta^- - i \theta^-{\bar       \partial}
\theta^0 +e^{-i\theta^0}{\bar     \partial  }\theta^+ (\theta^-)^2)
\nonumber\\
&&+ A^0(\frac{ 1   }{ 2   }     {\bar     \partial  }\theta^0+
i\theta ^- {\bar
\partial}  \theta^+e^{-i\theta^0})    \nonumber\\
&&
+A^-{\bar       \partial}\theta^+e^{-i\theta^0}
]\; .
\end{eqnarray}

As has been explicitly verified \cite{gerasimov}, this action is
nothing but the WZW action when $g$ is parameterized by the Gauss
decomposition (\ref{n.5}):
\begin{equation}
\label{5.37}
S =   S[g] + \frac{i}{\pi}{\rm Tr}(A g{\bar
\partial } g^{-1})    \; .
\end{equation}

Use of expression (\ref{n.4}) rather than (\ref{n.4a}) would have
resulted in the full gauged WZW model (\ref{gaugewzw}).  This may
readily be verified.  The $A {\bar A}$ terms which do not contain $g$
coupling originate from the $e^{i T^0 \theta^0}$ rotation.  The
general form of (\ref{5.31}) with $\bar A\neq 0$ follows by
substituting (\ref{5.22}) into (\ref{n.4}), which gives, after
performing the trace, additional terms like
\begin{equation}
\label{n.7}
\frac{1}{2 \pi}\int_0^1 d \epsilon \theta^0 {\bar A}^-
A^+ e^{\epsilon \theta^0}=\frac{1}{2\pi}(e^{\theta^0}-1)\bar A^- A^+ \; .
\end{equation}
Thus the epsilon integral gives an $A {\bar A}$ term without any
$\theta$ coupling.  It is therefore not necessary to view this term as
a counterterm that has to be added by hand to insure gauge invariance
\cite{polyakov,kaa}.  Since we are using a covariant form of the
infinitesimal anomaly (\ref{n.2}), the iterated anomaly is expected
and indeed found to be the gauge invariant expression
(\ref{gaugewzw}).

The bosonic Lagrangian (\ref{5.36}), or (\ref{5.37}), is equivalent to
that of the free fermionic model (\ref{5.17}), as originally
established by Witten \cite{witten}, and normally referred to as
non-abelian bosonization.  The result (\ref{5.36}) plays an essential
role in the path integral derivation of this equivalence \cite{tssg}.

Here we note that expression (\ref{5.36}) is also closely linked to
the Wakimoto \cite{wakimoto} or free field realization of Kac-Moody
algebras.  This correspondence is the field theoretical analogue of
the relation between the path integral for spin and the Dyson mapping
as has been discussed in ref. \cite{scholtz}.

To obtain the current-current generating functional in terms of
bosonic variables \cite{tssg} we have to integrate the action
(\ref{5.36}) with respect to the Haar measure which reads $
D\theta^+D\theta^0 D\theta^- \det( e^{-i\theta^0})$ in the Gauss
decomposition:
\begin{equation}\label{5.40c}
Z_{\rm B}[A]=\int \!D\theta^+\; D\theta^0 \;D\theta^- \det( e^{-i\theta^0})
e^{-k S}\; .
\end{equation}
Here we allowed for a slight generalization by introducing the
constant $k$.  For one fermion flavour, as we have here, $k=1$, but
for $k$ flavours the anomaly is yielded $k$ times so that the WZW
action occurs with a multiplicative constant $k$.  We also remark that
the fermion currents with $k$ flavours close on a Kac-Moody algebra of
level $k$.

As noted in \cite{gerasimov}, the expressions to which the $A$
currents couple in eq.  (\ref{5.36}) are closely related to the
Wakimoto realization \cite{wakimoto,dotsenko,felder}.  To make the
association explicit, we have to perform a change of variables
\begin{equation}
\label{5.41b}
\beta=k{\bar    \partial   } \theta^+e^{-i\theta^0}
\; .
\end{equation}
Naively the Jacobian of this transformation is
\begin{equation}
J_\beta = \frac{1}{\det(e^{-i \theta^0)}}
\frac{1}{\det( {\bar  \partial})}
\; ,
\end{equation}
in which case this determinant cancels the one from the invariant
group measure, with the result that we obtain a flat measure.
The generating functional  becomes
\begin{equation}\label{gfunc}
Z_{\rm B}[A]=\int \!D\beta\; D\theta^0  D\theta^-
\det(1/{\bar \partial})\;
e^{-S}\; .
\end{equation}
with
\begin{eqnarray}
\label{5.42}
S&=&\frac{  1  }{ \pi   }\int d^2 x
[\frac{  k  }{ 4   }
\partial \theta^0{\bar   \partial  }\theta^0 +
 \beta
\partial \theta^-
\nonumber
\\
&&- i A^+( ik  {\bar       \partial }\theta^- +k \theta^-{\bar
\partial} \theta^0 +    i\beta (\theta^-)^2)
\nonumber\\
&&-i A^0(\frac{ i k }{ 2   }     {\bar     \partial  }\theta^0-
\theta ^-
\beta ) \nonumber \\
&&-iA^-(i \beta)
]\; .
\end{eqnarray}
The change of variables (\ref{5.41b}) transforms the WZW action to
that of free bosonic fields (\ref{5.42}).  For this reason
(\ref{5.42}) is called a free field realization of the WZW model.  The
expressions to which the sources $A$ couple in (\ref{5.42}), may be
viewed as a kind of ``classical'' Wakimoto realization \cite{boer},
where corrections due to normal ordering have been neglected.  This
issue has partially been discussed in ref. \cite{gerasimov}, where it
was pointed out that there are corrections from the measure when we
perform transformation (\ref{5.41b}).  These corrections in particular
modify the coefficient of the $\partial \theta^0{\bar \partial
}\theta^0$ term.  Although a complete derivation is still lacking, one
expects similar corrections to the above classical Wakimoto currents
to obtain the correct expressions
\begin{eqnarray}
\label{5.53}
J^-&=& ik   {\bar       \partial }\theta^- +\sqrt{ 2   }\sqrt{    (k+2) }
\theta^-{\bar \partial} \theta^0 +    i\beta (\theta^-)^2
\nonumber\\
J^0 &=& \frac{ i \sqrt{    (k+2)} }{ \sqrt{   2 }   }     {\bar
\partial }\theta^0-
\theta ^-
\beta \nonumber \\
J^+&=&i \beta
\quad.
\end{eqnarray}
The charge at infinity may similarly be understood as originating from
such anomalous corrections in the path integral approach
\cite{gerasimov}.

In conclusion, we have presented a simple and systematic procedure to
evaluate finite chiral determinants, leading to a constructive
identification of the counterterm $A\bar A$. This procedure also
facilitates the identification of a free field or Wakimoto realization
of Kac-Moody algebras.  A constructive procedure to deal with the
quantum corrections to these currents is, however, still lacking, a
not uncommon situation when general changes of variables are performed
in the path integral setting.

\bigskip

{\bf Acknowledgements}: We acknowledge financial support by the
Foundation for Research Development.
N.T would like to thank the Theory Division at CERN for their warm
hospitality. H.B.G also acknowledges support by the Alexander von
Humboldt Stiftung and would like to thank Gottfried Holzwarth and Hans
Walliser for their hospitality at the University of Siegen.

\end{document}